\documentclass[12pt]{article}
\usepackage{epsfig,graphicx}

\title{Neutral 3-body system in a strong magnetic field: factorization and exact solutions}
\author{  Yu.A.Simonov,\\ Institute of Theoretical and Experimental
Physics\\ 117118, Moscow, B.Cheremushkinskaya 25, Russia}
\date{}

\newcommand{\be}{\begin{equation}}
\newcommand{\ee}{\end{equation}}

\def\fun#1#2{\lower3.6pt\vbox{\baselineskip0pt\lineskip.9pt
\ialign{$\mathsurround=0pt#1\hfil ##\hfil$\crcr#2\crcr\sim\crcr}}}

\newcommand{{\SD}}{\rm SD}

\newcommand{\ver}{\mbox{\boldmath${\rm r}$}}
\newcommand{\vesig}{\mbox{\boldmath${\rm \sigma}$}}

\newcommand{\veP}{\mbox{\boldmath${\rm P}$}}

\newcommand{\veq}{\mbox{\boldmath${\rm q}$}}

\newcommand{\vez}{\mbox{\boldmath${\rm z}$}}

\newcommand{\veL}{\mbox{\boldmath${\rm L}$}}

\newcommand{\veR}{\mbox{\boldmath${\rm R}$}}

\newcommand{\vexi}{\mbox{\boldmath${\rm \xi}$}}
\newcommand{\veta}{\mbox{\boldmath${\rm \eta}$}}
\newcommand{\veB}{\mbox{\boldmath${\rm B}$}}

\newcommand{\veJ}{\mbox{\boldmath${\rm J}$}}

\newcommand{\vepi}{\mbox{\boldmath${\rm \pi}$}}
\newcommand{\vemu}{\mbox{\boldmath${\rm \mu}$}}
\newcommand{{\Mc}}{\mathcal{M}}

\begin{document}

\maketitle
\begin{abstract}

Neutral systems containing two identical particles, in homogeneous magnetic field are shown
 to obey exact factorizable solutions both in nonrelativistic and relativistic
 formalism, similarly to the neutral two-body systems. Concrete examples of the
helium atom and the neutron as a (ddu) system are considered.

\end{abstract}

\section{}  The problem of composite system in the magnetic field has been always an
important topic of investigation and of textbooks. One particular, and
probably the simplest problem is that of the neutral two-particle system in MF, and
 a necessary step in its solution is the factorization of the c.m.
and internal motion, which in the external  MF is not a simple procedure. This
problem was solved in \cite{1,2,3,4} in the nonrelativistic framework.
Moreover, in \cite{4} a general theorem was given, stating existence of a set
of pseudomomenta in MF in a neutral $N$-body system. The factorization problem
in the relativistic context was recently solved in \cite{5}, where the
relativistic Hamiltonian for two-body neutral system was derived from the QCD
path integral and applied to find the neutral meson  spectrum in strong MF.

The idea of strong MF in our surroundings attracts nowadays a lot of
researchers and founds support and confirmation in many areas. In astrophysics
MF were known for a long time and very strong MF, up to $10^{18}$ Gauss, were
found in magnetars \cite{6}, very strong MF are possible in peripheral
heavy-ion collisions at RHIC
 and LHC \cite{7}, and  in early Universe \cite{8}.

On theoretical side the important development in last years was in the
study of hydrogene atom in strong MF \cite{9,10,11}, where it was shown \cite{9,10}, that
the spectrum of hydrogene atom is stabilized in the limit of high MF
due to $e^+e^-$ loop corrections to the Coulomb force.

The case of neutral three-body  system in MF is not less important, than its
two-body analog; however here the development is less active. In particular,
the general theorem of factorization of c.m. and internal motion found in the
case of two- body neutral system, is not known for the three-body case, and the
property of stabilization is not yet found.

The purpose of the paper is twofold. First, we present in section 2 the exact procedure
 of factorization in MF and demonstrate the nonrelativistic Hamiltonian for neutral three-body problem with two
 identical particles.

In section 3 this Hamiltonian is considered for the $^3He$ and $^4He$ atoms and
some properties of the spectra are discussed.  At the end of section 3  the
relativistic generalization
 of the same Hamiltonian is derived and the neutron or $\Delta^0$ isobar system is considered
 as physical examples, and dynamics and some properties of spectra are discussed. In section 5
 the results are summarized and prospectives are given.

\section{Three-body nonrelativistic Hamiltonian in the magnetic field}

The nonrelativistic (Pauli) Hamiltonian for  three particles with masses and
charges $m_i,e_i, i=1,2,3$ in the magnetic field $\veB$ has the standard form
\be H  = \sum_{i=1}^3 \frac{(p_k^{(i)} - e_i A_k)^2-e_i\vesig^{(1)} \veB}
{2m_i}\equiv H_0+H_\sigma \label{1}\ee

A general problem of few-body treatment in the magnetic field is the separation
of the c.m. and relative (internal)  motion. This problem is nontrivial and
allows a factorizable solution for neutral two-body system after a special
phase factor is introduced and conserved pseudomomenta are defined
\cite{1,2,3,4,5}. In \cite{5} this solution was generalized to the relativistic
two-body case. Below we show that the three-body system can be solved
(factorized) in the same way in one special
 case: when two of three particles are identical, i.e. $e_1=e_2, m_1=m_2$, but   $m_3$
is  arbitrary and $e_3 =- 2e_1$.

We define $e_1=e_2=-\frac{e}{2}, e_3=e$, $m_1=m_2=m$, and introduce Jacobi
coordinates\be \left\{ \begin{array}{l}
R_k=\frac{1}{m_+} \sum m_iz_k^{(i)},\\
\eta_k = \frac{z_k^{(2)} - z_k^{(1)}}{\sqrt{2}},\\
\xi_k =\sqrt{\frac{m_3}{2m_+}} (z_k^{(1)} + z_k^{(2)} - 2z_k^{(3)}).\end{array}
\right.\label{2}\ee where $m_+ =2m+m_3$.
Denoting
 \be \mathcal{P}_k\equiv
\frac{\partial}{i\partial R_k},~~ q_k \equiv \frac{
\partial}{i\partial \xi_k},~~ \pi_k \equiv \frac{\partial}{i\partial
\eta_k}\label{3}\ee one has \be p_k^{(i)}= \alpha_i  \mathcal{P}_k +\beta_i q_k
+ \gamma_i \pi_k,\label{4}\ee \be p_k^{(1)} = \frac{m}{m_+} \mathcal{P}_k
+\sqrt{\frac{m_3}{2m_+}} q_k- \frac{1}{\sqrt{2}}\pi_k\label{5}\ee

\be p_k^{(2)} = \frac{m}{m_+} \mathcal{P}_k +\sqrt{\frac{m_3}{2m_+}} q_k +
\frac{1}{\sqrt{2}}\pi_k\label{6}\ee \be   p_k^{(3)} = \frac{m_3}{m_+}
\mathcal{P}_k -\sqrt{\frac{2m_3}{m_+}} q_k \label{7}\ee

In terms of $P_k,q_k,\pi_k$ the Hamiltonian has  the form

$$ H_0= \frac{1}{2m} \left[ \frac{m}{m_+} \mathcal{\veP}
+\sqrt{\frac{m_3}{2m_+}}\veq -\frac{\vepi}{\sqrt{2}} +\frac{e}{4} \left(\veB
\times \left(\veR + \sqrt{\frac{m_3}{2m_+}} \vexi -
\frac{\veta}{\sqrt{2}}\right)\right)\right]^2+$$

$$+ \frac{1}{2m} \left[ \frac{m}{m_+} \mathcal{\veP}
+\sqrt{\frac{m_3}{2m_+}}\veq +\frac{\vepi}{\sqrt{2}} +\frac{e}{4} \left(\veB
\times \left(\veR + \sqrt{\frac{m_3}{2m_+}} \vexi +
\frac{\veta}{\sqrt{2}}\right)\right)\right]^2+$$
$$+ \frac{1}{2m_3} \left[ \frac{m_3}{m_+} \mathcal{\veP}
-\sqrt{\frac{2m_3}{m_+}}\veq -\frac{e}{{2}}\left(\veB \times \left(\veR
-\sqrt{\frac{2m^2}{ m_+m_3}} \vexi \right)\right)\right]^2$$ \be\equiv
\frac{1}{2m} \left((\veJ^{(1)})^2 + (\veJ^{(2)})^2\right)+\frac{1}{2m_3}
\veJ^{(3)}.\label{8}\ee

 We now do the same step as in the two-body case and introduce the phase factor, which
 in our case has the form
\be \Psi (\veR,\vexi, \veta) = e^{-i\frac{e}{4} (\veB\times \veR) \vexi
\sqrt{\frac{2m_+}{m_3}} +i\veP\veR} \varphi (\xi,\eta)\equiv e^{i\Gamma}
\varphi\label{9}\ee Acting with operators $\veJ^{(i)}$ on $\Psi$ in the form of
(\ref{9}), one obtains a remarkable simplification,

 \be H_0\Psi = H_0 e^{i\Gamma} \varphi = e^{i\Gamma} \tilde H_0 \varphi,\label{10} \ee where  e.g.
for $\veP=0$ one has
$$ \tilde H_0 =- \frac{1}{2m} (\Delta_\xi + \Delta_\eta) +
\frac{1}{2m} \left(\frac{eB}{4}\right)^2 \left( \frac{m^2_+}{m^2_3}
(\vexi_\bot)^2+ (\veta_\bot)^2\right)+$$ \be +\frac{eB_k}{4m} \left( \frac{m_3
-2m}{m_3} L_k^{(\xi)} + L^{(\eta)}_k\right)\label{11}\ee Here $L_k^{(\xi)},
L_k^{(\eta)}$ are Jacobi angular momenta \be\veL^{(\xi)}=\left (\vexi\times
\frac{\partial}{i\partial \vexi}\right), ~~ \veL^{(\eta)} = \left( \veta \times
\frac{\partial}{i\partial \veta}\right)\label{12}\ee One can see in (\ref{11}),
that eigenfunctions of $\tilde H_0$ factorize,
 \be
 \varphi(\vexi,\veta) = f_1 (\vexi_\bot) f_2(\veta_\bot) \exp (ik_\xi \xi_3) + i k_\eta  \eta_3)\label{13}\ee
where $f_1, f_2$ have standard form, e.g.

\be f_1 (\veta_\bot) = \frac{e^{il_\xi \varphi_\xi}}{\sqrt{2\pi}} \chi_n (x),
~~ l_\xi=0,\pm 1,... \label{14}\ee \be \chi_n (x) = C_n e^{-\frac{x}{2}}
x^{\frac{|l_\xi|}{2}} F(-n_\xi, |l_\xi|+1, x),\label{15}\ee
 where $n_\xi =0,1,2,..., C_n$ is the normalization
constant, and $$ x=\frac{eBm_+}{4m_3} \vexi^2_\bot, $$ while the corresponding
energy is \be E_{n_\xi} = \frac{k^2_\xi}{2m} + \frac{eBm_+}{2mm_3} \left(n_\xi+
\frac{1+|l_\xi|}{2} + \frac{m_3-2m}{2m_3} l_\xi\right).\label{16}\ee
 In the
same way for $f_2(\veta)$ one obtains \be f_2(\veta_\bot) \equiv \frac{\exp
(il_\eta\varphi_\eta)}{\sqrt{2\pi}} \chi_{n_\eta} (y), ~~ l_\eta =0, \pm
1,...\label{17}\ee \be \chi_{n_\eta}(y) =C_{n_\eta }e^{-\frac{y}{2} }
y^{|l_\eta|/2} F(-n_\eta, |l_\eta| +1, y),\label{18}\ee where
$y=\frac{eB}{4}\veta^2_\bot, n_\eta =0,1,2,...$ and the energy of the
$\eta$--motion is \be E_{n_\eta} = \frac{k^2_\eta}{2m} + \frac{eB}{2m} \left(
n_\eta + \frac{1+ |l_\eta|+l_\eta}{2}\right)\label{19}\ee

The total energy of $(H_0+H_\sigma)$ is

\be E=E_{n_\xi}+E_{n_\eta} - \sum^3_{i=1}  \frac{e_i\vesig^{(i)}
\veB}{2m_i}\label{20}\ee

\section{Physical examples}

a)  The case of Helium atom.

As a first example we consider  the neutral atomic system of a helium atom,
where two electrons play the role of identical particles,
$m_1=m_2=m_e,e_1=e_2=-e, e_3=2e$, and $m_3$ is the mass of the helium nucleus,
$m_3=M(^3He)$ or $m_3=M(^4He)$. In the first case for the ground state atom the
electrons on the $S$ level have opposite spin directions, and $H_\sigma$ in
(\ref{1}) reduces to the magnetic moment term of $^3He$ nucleus,
$H_\sigma=-\vemu(^3He )\veB$. In the case of $^4He$ the term $H_\sigma$ in
(\ref{1}) for the ground state is identically zero.

The Coulomb interaction $$ V_{\rm Coul} (|\vez_i - \vez_j|) =-
\frac{\alpha}{|\vez_i-\vez_j|},
$$ is introduced in the standard way, adding to $H_0+H_\sigma$ in (\ref{1}) the term
\be V_{\rm Coul}^{He} (\xi,\veta) = 2V_{\rm Coul} \left( \left|
\sqrt{\frac{m_+}{2m_3}}\vexi- \frac{\veta}{\sqrt{2}}\right|\right) + 2 V_{\rm
Coul} \left(\left|\sqrt{\frac{m_+}{2m_3}}\vexi+
\frac{\veta}{\sqrt{2}}\right|\right)- V_{\rm Coul} (|\sqrt{2} \veta
|).\label{21}\ee

In absence of MF, the Hamiltonian for the Helium atom is usually written in the
form, which neglects finite nucleus mass corrections, namely \be
h=-\frac{1}{2m} (\Delta_1+\Delta_2) - 2\alpha \left( \frac{1}{r_1}+
\frac{1}{r_2}\right) + \frac{\alpha}{r_{12}}; ~~ \ver_i =\vez_i -\vez_3, ~~
i=1,2,\label{24}\ee while our Hamiltonian (\ref{11}) for $\veP=0$ contains
those (cf connection of $\veta,\vexi$ and $\ver_i$ in (\ref{2})). \be
H=-\frac{1}{2m} (\Delta_\xi +\Delta_\eta) + V_{\rm Coul}^{He} (\vexi, \veta) =h
-\frac{1}{2m_3} \left(\frac{\partial}{\partial\ver_1}+
\frac{\partial}{\partial\ver_2}\right)^2.\label{25}
 \ee

Accurate calculations with the Hamiltonian (\ref{24}) were being done for a
long time  \cite{12} and have achieved an extremely high level of accuracy, see
e.g. \cite{13}(see \cite{14} for a recent review).

When strong MF is present, one can make the adiabatic approximation, as in  the
hydrogene atom
 case
\be H_{\rm adiab} =-\frac{1}{2m_3} \left(\frac{\partial^2}{\partial\xi^2_3}+
\frac{\partial^2}{\partial\eta^2_3}\right)+ V_{\rm adiab} (\xi_3,
\eta_3),\label{26}\label{25} \ee where $V_{\rm adiab}$ is \be V_{\rm
adiab}(\xi_3,\eta_3) = \int V_{\rm Coul}^{He} (\vexi, \veta) d^2\xi_\bot
d^2\eta_\bot f_1^2(\xi_\bot) f^2_2(\eta_\bot)\label{27} \ee

As a result, the problem reduces to the one-dimensional three-body problem with
Coulomb-like interaction. Neglecting c.m. corrections and the repulsive $ee$
interaction term, one can factorize the wave function $$\psi(\xi,\eta) \to
\psi(r_1)\psi(r_2)
$$
and in the adiabatic approximation one has a product of one-dimensional
hydrogene-like functions, which obey stabilized dynamics at large
 MF \cite{9,10, 11}. This approach can be generalized in the same way, as it is done for the
Helium atom without MF \cite{12,13,14}. Since the $ee$ interaction is repulsive
one can establish a lower bound for the Helium binding energy in MF as the
twice the  limiting binding energy of the  hydrogene atom in MF, i.e. $2\cdot
1.74$ keV=3.48 keV.

b) Neutral baryon case.

For neutron or $\Delta^0$ baryon with the structure (ddu), of $\Xi^0$ baryon,
(ssu), one can use the same strategy, as in the nonrelativistic case, but one
must replace the starting form (\ref{1}) by its relativistic analog (see
\cite{5}
 and refs. therein for details). The simple replacement holds in the external magnetic
field, when one can keep, as in (\ref{1}), the $(2\times 2)$ structure,
neglecting connection to the Dirac underground.

In this case the new Hamiltonian reads \be H_{\rm rel}=(H_0+H_\sigma) (m_i\to
\omega_i) + \frac{m_1^2+\omega^2_i}{2\omega_i} + W,\label{30} \ee where $W$
contains gluon exchange (color Coulomb) $V_{GE}$, confinement $V_{\rm conf}$ ,
and spin-dependent and selfenergy terms.

The eigenvalue of (\ref{30}), $M(\omega_1, \omega_2, \omega_3)$, is subject to
the stationary point conditions $\left.\frac{\partial M}{\partial
\omega_i}\right|_{\omega_i=\omega_i^{(0)}}=0$, which  define $\omega_i^{(0)}$
and the final eigenvalue $M(\omega_1^{(0)}, \omega_2^{(0)}, \omega_3^{(0)}).$
As in the nonrelativistic case, one finds the full separability of the
Hamiltonian $H_{rel}$ in case, when
$\omega_1^{(0)}=\omega_2^{(0)}=\omega^{(0)}$. This is possible when not only
 $e_1=e_2, m_1=m_2$, but also spin projections of both quarks are equal, $\vesig_1 \veB=\vesig_2\veB$.
 For this configuration one can calculate baryon masses as functions of $B$, as it was done
 in \cite{5} for mesons.

We leave this topic for another publication.

\section{Summary and conclusions}

We have  found a factorizable and fully separable form of nonrelativistic and
relativistic Hamiltonian for
 a neutral three-body problem in MF with two identical particles.

We have found in this case exact solutions, and  consider the physical examples
of helium atom and neutral baryon. We demonstrate, that the Coulomb or  gluon
exchange attraction at large MF can  cause the problem of stability as in the
case of hydrogene
 atom, and for the helium atom the stability is ensured by that of the hydrogene atom. Our formalism may pave the road for the accurate calculations
of three body system in MF.
The author is grateful to M.A.Andreichikov and B.O.Kerbikov for discussions.

\end{document}